# Nanochannels for Spin-Wave Manipulation in $Ni_{80}Fe_{20}$ Nanodot Arrays

Sourav Sahoo[1], Surya Narayan Panda[1], Saswati Barman[2], Yoshichika Otani[3,4] and Anjan Barman[1,*]

[1]*Department of Condensed Matter Physics and Material Sciences, S. N. Bose National Centre for Basic Sciences, Block JD, Sector III, Salt Lake, Kolkata 700106, India*

[2]*Institute of Engineering and Management, Sector V, Salt Lake, Kolkata 700091, India*

[3]*CEMS-RIKEN, 2-1 Hirosawa, Wako, Saitama 351-0198, Japan*

[4]*Institute for Solid State Physics, University of Tokyo, 5-1-5 Kashiwanoha, Kashiwa, Chiba 277-8581, Japan*

*\*Email address: abarman@bose.res.in*

## Abstract

Patterned magnetic nanostructures are potential candidates for future energy efficient, on-chip communication devices. Here, we have experimentally and numerically studied the role of nanochannels to manipulate spin waves in $Ni_{80}Fe_{20}$ connected nanodot arrays of varying filling fraction. Rich spin-wave spectra are observed in these samples, where the number of spin-wave modes decreases with increasing filling fraction due to the retrenchment of the demagnetizing field. The nanochannels affect the spin-wave modes of the connected dots through dipole-exchange coupling. For all modes the vertical nanochannels couple the nanodots, except for the highest frequency modes where all nanochannels act as coupler. This feature is further explored in the simulation, which reveals that only the highest frequency mode can propagate through all the nanochannels, analogues to an electronic demultiplexer. This study will be useful to understand the role of nanochannels in patterned magnetic nanostructures and their applications in spin-wave based communication devices.

Keywords: Magnonic crystal, Time-resolved Kerr microscopy, Spin wave, Nanochannel, Demultiplexer, Micromagnetic simulation.

# 1. Introduction

Periodically patterned thin film magnetic nanostructures commonly known as magnonic crystal (MC), is an interesting research field since its inception in 1996 [1] due to its potential applications in wave-based computing, on-chip GHz frequency data communication, processing, low energy nanoelectronic devices among others. Spin wave (SW) is the propagation of perturbed energy from the equilibrium state energy in the form of phase in a magnetically coupled system, which acts as information carrier in magnonics [2,3]. In addition to the fundamental understanding of SW in various one-, two- and three-dimensional MCs [4], a number of miniaturized components and devices such as SW-based multiplexer [5], interferometer [6], grating [7], waveguides [8,9], phase shifters [10], directional couplers [11], nanomagnetic antenna [12] and neuromorphic computing [13] have been demonstrated during last one decade.  More recently strong coupling of magnons with microwave photon [14], phonon [15], magnon [16,17] and superconducting qubits [18] have been explored as hybrid systems for quantum transduction as well energy harnessing from alternative sources. Control of magnon propagation using topological spin texture [19] has landed a new research field of spin-texture controlled magnonics.

In essence, complete understanding and control of SWs in patterned arrays of nanomagnets is at the core of the research in magnonics. The overwhelming progress in nanofabrication and high-frequency measurement techniques have made it possible to explore the SW dynamics in various complex patterned magnetic nanostructures. In the last decade, there have been a large number of study of SW propagation, localization, hybridization and damping on different types of patterned structures such as physically isolated periodically patterned magnetic structures, namely dot lattices [20-26], nanowire and nanostripes [27-29] and connected structures, namely  antidot (AD) lattices [30-36]. The main focus of studying patterned magnetic nanostructures is to understand and tune the SW frequency, nature of modes, mode profiles and damping by varying the element width [21,23,28,37], shape [26,34,35], lattice constant [24,38], lattice symmetry [25,36], base material [35,39] as well as by changing the external bias magnetic field strength and orientation [32-34]. Further, the filled AD lattices [40] and other form of bicomponent MCs [41] offer more freedom to tune the SW properties. There have been a large number of studies on the SW dynamics in ferromagnetic AD lattices. Most of the studies were focused on the variation of diameter, shape, separation and lattice symmetry of ADs, which dealt with magnetic elements connected by irregular shaped nanochannels (NCs). The complex demagnetizing field profiles in such structures made it non-trivial to understand the role of NCs on the SW spectra in these systems.

Here, we have investigated the SW dynamics in two-dimensional (2D) square arrays of $Ni_{80}Fe_{20}$ (Permalloy; Py hereafter) nanodots which are physically connected to its neighbouring nanodots by simple rectangular shaped NCs of the same material. We have varied the dot size ($S$) and the NC length ($l$), while keeping the NC width ($W$) constant. The filling fraction (area covered by Py/area of one unit cell) increases with decreasing $S$ and $l$. We have exploited a custom-built time-resolved magneto-optical Kerr effect (TR-MOKE) microscope to reliably measure the local precessional dynamics of the samples. We have observed rich SW spectra in these connected nanodot (CND) arrays. The number of SW modes decreases with increasing filling fraction due to the reduction in the demagnetizing region. All the SW modes are found to be coupled between the dots through the vertical NC (VNC), connecting the nanodots perpendicular to applied magnetic field ($H$) except the highest frequency mode, which shows a mixed backward volume-Damon Eshbach (BV-DE) behaviour. On the contrary, the power of the mode is mainly concentrated inside the horizontal NC (HNC), connecting the nanodots parallel to $H$. Further, using numerical micromagnetic simulations we have shown the role of NCs to modulate the higher frequency modes due to the dipole-exchange coupling between the dots.

## 2. Experimental Details

The square shaped arrays Py CNDs each with a 25 × 25 µm$^2$ area were fabricated by using a combination of electron-beam lithography, electron-beam evaporation and ion milling on self-oxidized Si (100) substrate. The 20-nm-thick Py film coated with a 60-nm-thick Al$_2$O$_3$ protective layer was deposited in an ultra-high vacuum chamber at a base pressure of 2 × 10$^{-8}$ Torr on a commercially available self-oxidized Si(100) substrate. Bi-layer poly(methyl methacrylate)/methyl methacrylate (PMMA/MMA) resist was used for e-beam lithography to make the resist pattern on the Py thin film followed by ion milling at a base pressure of 1 × 10$^{-4}$ Torr with a beam current of 60 mA for 6 min for creating the pattern.

The ultrafast magnetization dynamics of the connected nanodot arrays was measured by using an all-optical TR-MOKE microscope based on a two-colour collinear pump-probe technique[42]. The second harmonic (λ = 400 nm, pulse-width = 100 fs, fluence = 18 mJ/cm$^2$) of a Ti-Sapphire oscillator (Tsunami, Spectra-Physics) was used to excite the dynamics, while the time-delayed fundamental beam (λ = 800 nm, pulse-width = 80 fs, fluence = 2 mJ/cm$^2$) was used to detect the ensuing Kerr rotation as a function of the delay time. The time resolution of the system is about 100 fs which is determined by the cross-correlation between the pump and the probe beam. The collinear pump-probe beam is focused on the sample by a single microscope objective (MO) of numerical aperture, N.A. = 0.65. The probe beam is tightly focused with ~800 nm spot size on the sample surface by the MO, where the pump beam becomes slightly defocused with about 1 µm spot size, and they are precisely overlapped and placed on the desired region of the sample by using a piezoelectric x-y-z scanning stage. This ensures the measurement of the magnetization dynamics form uniformly excited region of the sample. The TR-MOKE microscope is equipped with a feedback loop of the scanning stage, a CCD camera and a white light illumination system for better stability and viewing. The same MO collects the back-reflected beams, out of which the pump beam is filtered out and the probe is steered to an optical bridge detector to simultaneously measure the time-resolved Kerr rotation and reflectivity. Two lock-in amplifiers are used to measure the signals separately in a phase sensitive manner by using the electrical output from an optical chopper, which is used to chop the pump beam at 2 kHz frequency. A static magnetic field is applied at a small angle of about 10° from the sample plane which generates a small out-of-plane demagnetizing field. The in plane component of this magnetic field is defined as the bias field, $H$. After the laser induced ultrafast demagnetization, the out-of-plane component of demagnetizing field is modified rapidly, inducing a damped precessional motion in the sample.

## 3. Results and Discussions

### 3.1. Experimental Results

Figure 1(a) shows the scanning electron micrographs of the CNDs with variable $S(l)$ of 865 nm(320 nm), 660 nm(210 nm) and 465 nm(105 nm) nm, denoted as S1, S2 and S3, while $w$ is constant at around 260 nm for all samples. Up to ± 10 nm deviation is observed in the lateral dimensions of the samples. The filling fraction is calculated for all samples and tabulated in Table I. For S3, the edges of nanodots and NCs are slightly deformed. These deviations and deformations have been corroborated in the micromagnetic simulation.

Table I – Filling factor ($f_f$) for three sample is calculated by taking the ratio of the area covered by material (Py) within a unit cell with the area of unit cell of square. $f_f = A_{py}/D^2$; $(D = l + S)$.

| Sample | Dot Size ($S$ nm) | Connector length ($l$ nm) | Filling factor ($A_{py}/A$) |
|---|---|---|---|
| S1 | 865 | 320 | 0.64 |
| S2 | 660 | 210 | 0.72 |
| S3 | 465 | 105 | 0.83 |

The experimental geometry for the TR-MOKE measurement is shown in Fig. 1(b). Typical time-resolved reflectivity and Kerr rotational data from S1 at $H = 1.67$ kOe are shown in Fig.

1(c) and 1(d), respectively. A schematic of sample geometry is shown in the inset of Fig. 1(c). The time scale is broken between 3 to 10 ps to highlight the ultrafast magnetization dynamics[42,43] in different temporal regimes. Following the zero delay, the Kerr rotation exhibits a sharp drop due to the laser induced ultrafast demagnetization (Region II), a fast magnetization recovery or fast remagnetization (Region III) and a slow remagnetization (Region IV) superimposed with multimodal damped precessional motion. The ultrafast demagnetization and fast remagnetization data are fitted with an expression derived from three-temperature model [44], which yields demagnetization time of ~200 fs and fast remagnetization time of ~1 ps. The slow remagnetization time is obtained as ~200 ps by fitting the data with an exponential function. The background subtracted time-resolved Kerr rotation data for all three samples at $H = 1.67$ kOe is shown in Fig. 2(a). The fast Fourier transformation (FFT) spectra of time-resolved data yield the frequencies of the SW modes (Fig. 2(b)). For S1, highest number of SW mode are observed and the mode number decreases with increasing filling fraction. The decrease in number of modes is due to the decrease in the demagnetized region with increasing filling fraction from S1 to S3. The lower frequency modes are suppressed as we move from S1 to S3 and consequently the SW frequency narrows down from 6.9 GHz to 3.6 GHz.

## 3.2. Micromagnetic Simulations

We have investigated the origin of the SW modes by numerical calculation using Object Oriented Micromagnetic Framework (OOMMF) software [45] considering 2D periodic boundary condition to mimic the large sample area. The arrays was divided into rectangular prism-like cells with size $4\times4\times20$ nm$^3$. The lateral cell size is well below the exchange length of Py ($\approx 5.2$ nm). The observed deformation in shapes was introduced in the simulations and the material parameters were used in the simulation as gyromagnetic ratio $\gamma = 17.6$ MHz/Oe, saturation magnetization $M_s = 860$ emu/cc, anisotropy field $H_K = 0$ and exchange stiffness constant $A_{ex} = 13 \times 10^{-6}$ erg/cm. The material parameters were extracted by fitting the measured bias field dependent precessional frequency of an unpatterned thin film with Kittel formula [46] (shown in Fig. S1 of the Supplementary Materials). The value of $A_{ex}$ is obtained from the literature [47]. The equilibrium magnetic states were obtained by relaxing the sample under study at a bias field of 1.67 kOe for sufficiently long time after reducing it from a large magnetic field so that the torque on the system goes below $10^{-6}$ A/m. The damping parameter is artificially made very large at $\alpha = 0.99$ in order to reach equilibrium quickly. For simulation of magnetization dynamics, the pump beam in the experiment was mimicked by applying a tickle field with amplitude of 20 Oe along z-direction with rise and fall time of both 10 ps and duration of 20 ps, while damping constant $\alpha = 0.008$ was used from the literature [48]. The FFT of simulated time-resolved out-of-plane magnetization component ($m_z$) is shown in Fig. 2(c), revealing the SW modes. The experimentally observed SW modes are qualitatively reproduced in the simulation. The experimental modes can be identified and their frequency spacing is reasonably well reproduced in the simulation. However, the relative powers of the modes are not always reproduced in the simulation. For example, in S1(S2), mode 5(4) is the highest intensity mode in both experiment and simulation, whereas in S3, relative powers of the modes do not agree. This is most likely due to the increasing deviation of the actual shape of the sample from the ideal shape, which is not possible to be fully reproduced in the finite difference method based OOMMF simulation. To understand the nature of the SW modes, we have further simulated the SW mode profiles using a home-built code [49]. The simulated power and phase maps of the SW modes for all three samples are shown in Fig. 3. The mode 1 (M1) of S1 and S2 is found to be an edge mode (EM), where the power of SW is concentrated mainly within the edges of VNCs and nanodots. The mode M2 of S1, S2 and M1 of S3 show extended behaviour in a DE-like geometry, i.e. the power of SW mode is extended along VNCs. The M3, M4 and M5 of S1 all show quantized behaviour in the BV-like geometry, i.e. with quantization axis parallel to the bias magnetic field direction with quantization number $n = 5$, 7 and 13, respectively. The M3 of S2 and M2 of S3 also show quantized behaviour in BV-like geometry with $n = 5$ and 7, respectively. The higher frequency modes, i.e. M6 and M7 of S1, M5 and M6 of S2 and M3 of S3 show mixed quantized behaviour in the DE-BV-like geometry. Noticeably here the SW modes are coupled between the nanodots through the VNCs except for the highest frequency modes where modes are coupled through both VNCs and HNCs. It is clear from the power profile that the power of the highest frequency mode (M7) of S1 is concentrated within the HNCs and shows uniform behaviour within HNCs. This behaviour is suppressed for the highest frequency mode of S1 and S2. The SW phase profile of all modes is shown at the top right corner of power maps. The color bars of power and phase profiles are presented at bottom right corner. To get a deep insight on the role of NCs in SW mode behaviour, we have performed some additional numerical calculations.

We have calculated magnetostatic field distribution within the nanodots and NCs using LLG micromagnetic simulator [50]. The magnetostatic field distribution within the CNDs is shown in Figs. 4 (a) for all three samples. Figures 4 (b) and (c) show the line scans of the internal magnetic field ($B_{in}$) map within the CNDs (marked as dotted lines R1 and R2 in Fig. 4(a)) along x-direction for all three samples. The inset of Figs. 4 (b) and (c) show the variation of $B_{in}$ within the HNC and VNC, respectively. The distribution of $B_{in}$ within HNC of S1 shows a central dip, which might be a reason for the concentrated power of the highest frequency mode within HNC of S1. The values of $B_{in}$ within HNC and VNC are plotted as a function of filling fraction in Fig. 4(d). With increasing filling fraction, $B_{in}$ also increases indicating a decrease in the demagnetizing field. The number of SW modes decreases with increasing filling fraction due to the decrement of exchange decoupled spins at the edges of the holes which suppress the edge modes. The value of $B_{in}$ is always higher in HNC than VNC as the bias field is applied along the HNC-direction. Hence, only the higher frequency modes are stabilized within the HNCs.

The difference in $B_{in}$ between HNC and VNC is shown in the inset of Fig. 4(d). With the increasing filling fraction this difference decrease (it is maximum in S1), which might be another reason that the power of highest frequency SW mode is concentrated within HNC. To explore the origin of the highest frequency modes, we have simulated the magnetization dynamics of physically isolated nanodots and NCs. The FFT spectra of the simulated time-resolved magnetization of the isolated nanodots and NCs are shown in Fig. 5(a). We observe three SW modes for nanodots and four SW modes for NCs. For isolated dots, the most intense peak shows uniform precession over the dot area. On the other hand, for the connector-only array two intense peaks are observed, which are uniform modes in the VNC (lower frequency) and HNC (higher frequency). In the CND array, the connector modes are suppressed while the nanodot modes become dominant indicating the sacrificial roles of the connectors only to boost the SW propagation between the nanodots.

We have further investigated the propagating nature of M3, M5 and M7 of S1 as these modes show distinct behaviour as discussed earlier, which are shown in Figs. 6(a) - (c). The SW modes are excited by a sinusoidal point excitation of desired frequency at the centre of the CND array as marked by 'I' in Fig. (6). Here, M3 and M5 propagate only through the VNCs along the vertical directions, whereas M7 propagates along all directions through all four NCs from the point of excitation and we obtain outputs at all four points as marked by 'O' in Fig. 6(c). This behaviour draws an analogy to electrical one-to-four demultiplexer logic operations. This remarkable observation of frequency selective SW propagation through nanochannels and demultiplexing-like operation for this special type of nanomagnetic array would be useful for the development of wave-based communications and computing devices.

## 4. Summary

In summary, we have investigated the role of magnetic nanochannels to control the SW properties in CND arrays. We have experimentally and numerically studied the time-resolved magnetization dynamics in the CND arrays and explored the effects of NCs connecting the nanodots on the SW properties through dipole-exchange interaction. The SW properties are found to vary strongly with the length of NCs. With the increase in filling fraction of the array the nonuniform demagnetized regions decreases letting the magnetostatic field to increase. Consequently, the number of SW modes also decreases with the increase in filling fraction. The experimental results are qualitatively reproduced by micromagnetic simulations, the latter further aids to identify various SW modes, namely the edge mode, extended mode, quantized mode and mixed quantized mode in the CNDs. The VNCs are found to couple the SW modes between the nanodots, except for the highest frequency mode, for which both NCs act as coupler. Simulation further reveals that the highest frequency SW mode primarily originates from the HNCs. Additional simulation reveals that the extended mode propagates only through the VNCs, whereas the highest frequency quantized mode propagates through all NCs, analogous to a four-way demultiplexer, which can be useful for future SW-based communication devices. This study will be helpful for the use of nanochannels in future magnonic devices.

**Acknowledgements:** AB gratefully acknowledges the financial support from S. N. Bose National Centre for Basic Sciences, India (Grant No. SNB/AB/18-19/211). SS and SNP acknowledge S. N. Bose National Centre for Basic Sciences for senior research fellowship. SB acknowledges Science and Engineering Research Board (SERB), Grant No. CRG/2018/002080 for funding. We acknowledge the technical help of Ruma Mandal for sample preparation and D. Kumar for some of the numerical simulations.

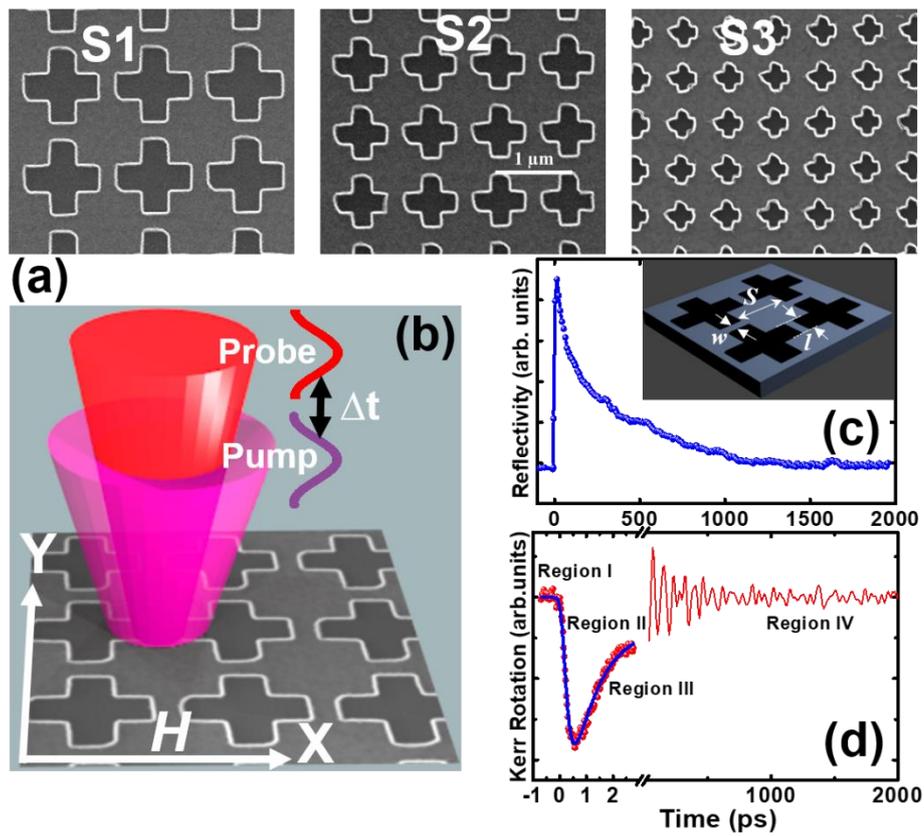

**Fig. 1.** (a) Scanning electron micrographs for three sample S1, S2 and S3. (b) Schematic of experimental geometry is shown. The configurations of of pump and probe beams and direction of applied magnetic field are highlighted. Typical time resolved (c) reflectivity and (d) Kerr rotation data of sample S1 are shown for $H$ = 1.67 kOe. At the inset of (c) a schematic of sample is shown where dot size, nanochannel length and width are marked.

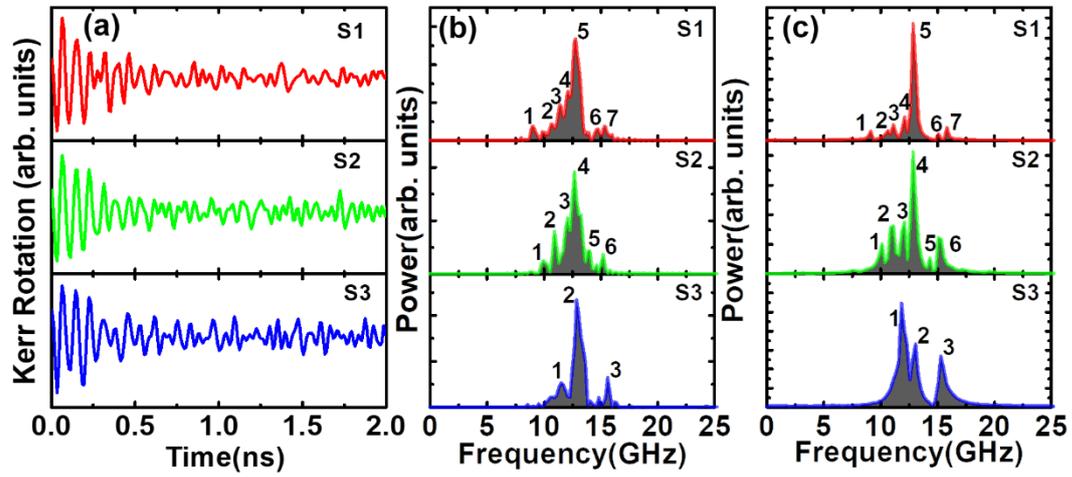

**Fig. 2.** (a) Experimental time-resolved Kerr rotation showing precessional oscillation for all three samples at $H = 1.67$ kOe. The FFT power spectra of time-resolved (b) experimental and (c) simulated precessional oscillation are shown. SW modes are numbered in the ascending value of frequency.

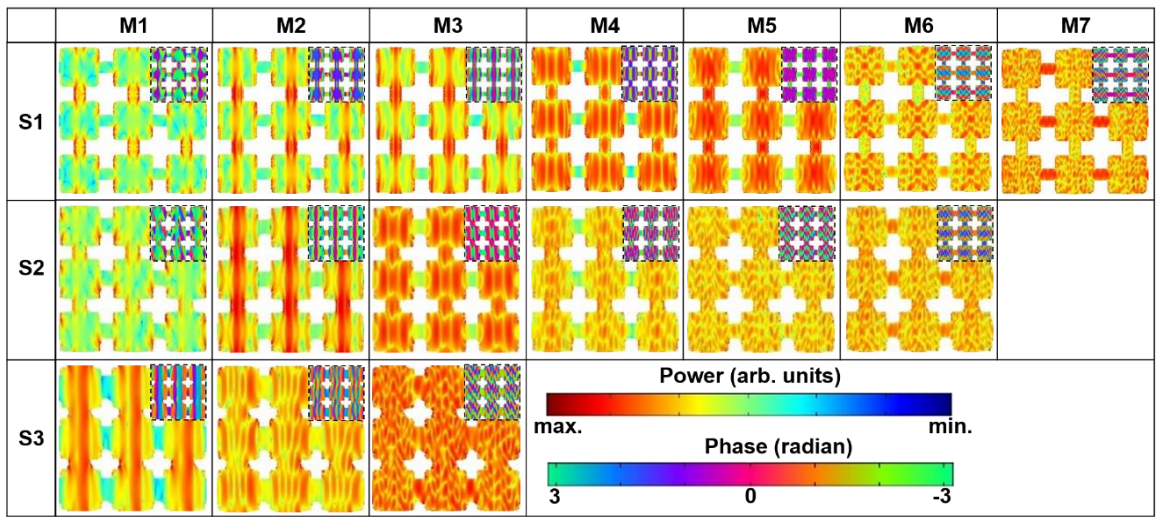

**Fig. 3.** Simulated spatial distributions of power of SW modes are shown. The corresponding phase profiles are shown in the inset.

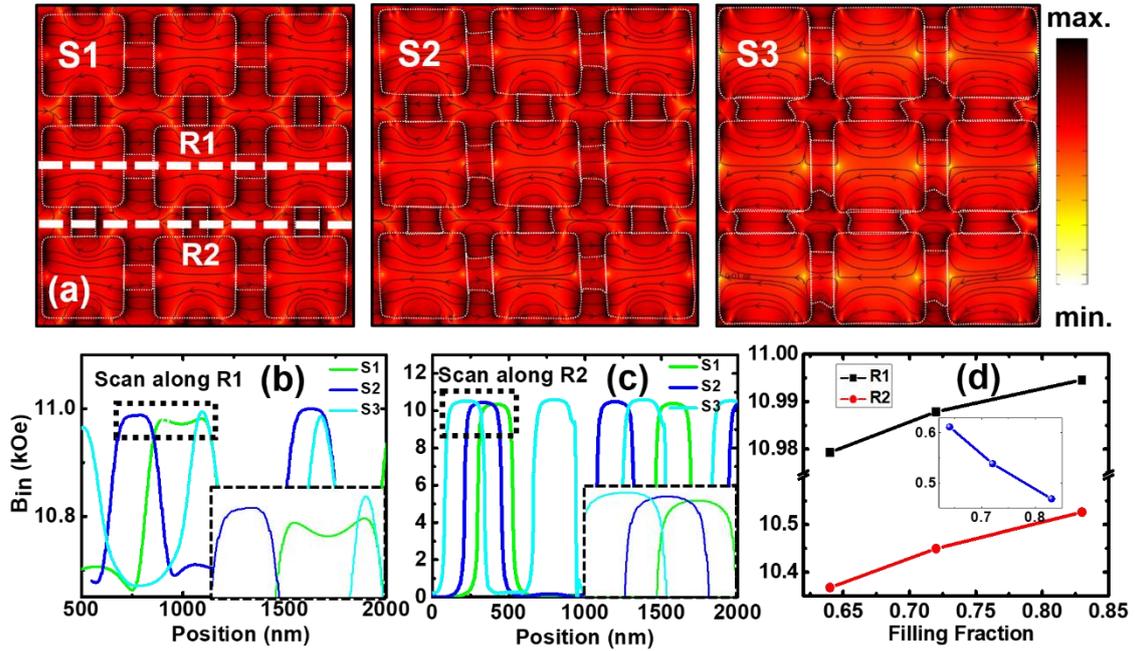

**Fig. 4.** (a) Magnetostatic field distribution of all the samples at $H$ = 1.67 kOe . Linescans of the magnetostatic field distributions taken along (b) horizontal nanochannels (R1) and (c) vertical nanochannels (R2) as shown by dotted lines on S1 of (a). (d) Variation of magnetostatic field with filling fraction. The difference in $B_{in}$ between HNC and VNC is shown in the inset.

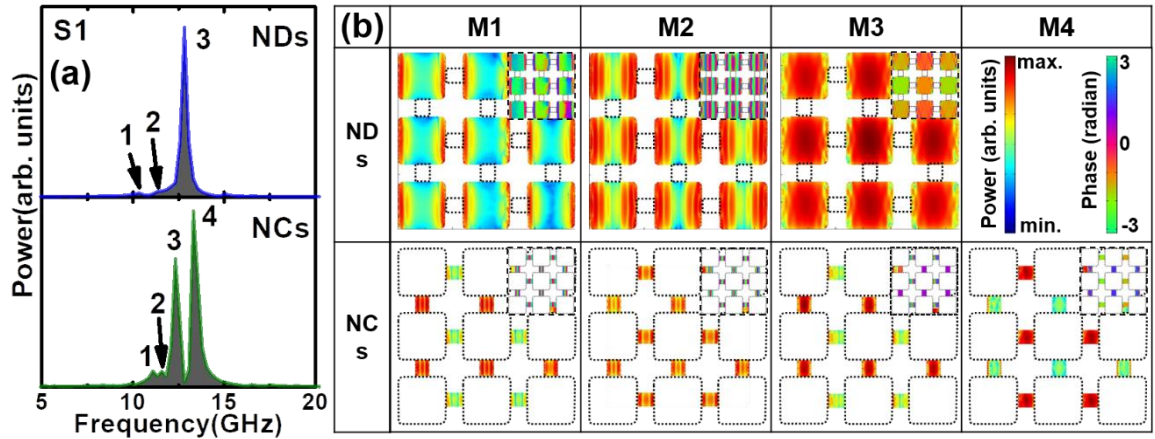

**Fig. 5.** (a) Simulated of power spectra of only nanodots and nanochannels of S1 are shown. Modes are marked by numbers in order of ascending frequency value. (b) Simulated spatial distribution of power of SW modes are shown. The corresponding phase profile is presented at the top right corner of each power map. The colour maps of power and phase profiles are shown at top right side of (b). Black dashed lines are to guide to eye to show the positions of nanodots or nanochannels.

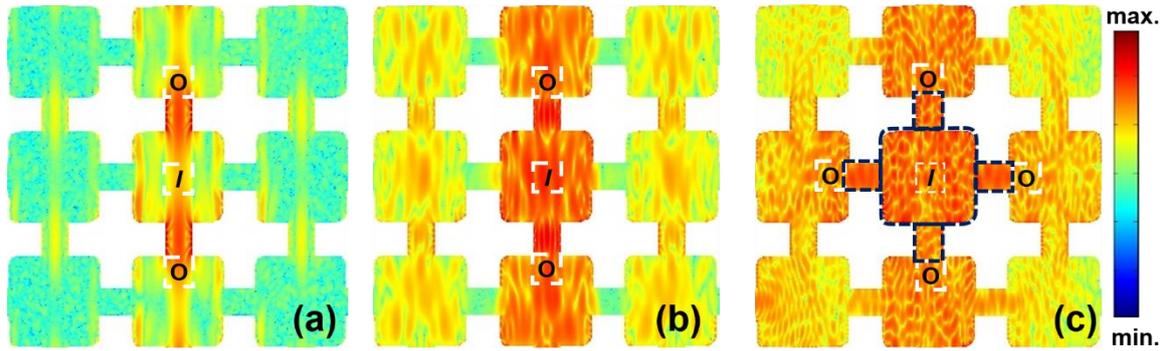

**Fig. 6.** Simulations showing propagating nature of spin-wave modes (a) M3, (b) M5 and (c)M7. Spin wave of desired frequency is excited at the centre (marked by *I*) of the array highlighted by dark blue coloured dashed line as shown in (c). Here '*O*' corresponds to the output of the spin-wave propagation through the nanochannels. M7 shows a demultiplexer-like functionality. The colour map of spin-wave power is shown at the right side.